\documentclass[a4paper, twocolumn]{revtex4-2} 
\usepackage{color}
\usepackage{graphics}
\usepackage{epsfig}
\begin{document}
	
	\title{Muon spin relaxation and emergence of disorder-induced unconventional dynamic magnetic fluctuations in Dy$_{2}$Zr$_{2}$O$_{7}$}
	
	\author{Sheetal$^{1,\dagger}$, Pabitra K. Biswas$^{2,\$}$, K. Yokoyama$^{2}$, D. T. Adroja$^{2}$ and C. S. Yadav$^{1*}$}
	\affiliation{$^{1}$School of Physical Sciences, Indian Institute of Technology Mandi, Mandi-175075 (H.P.), India\\
	$^{2}$ISIS Pulsed Neutron and Muon Source, STFC Rutherford Appleton Laboratory, Harwell Campus, Didcot, Oxfordshire OX11 0QX, United Kingdom}
	\altaffiliation{$^{\dagger}$ Current Address: Jülich Centre for Neutron Science JCNS at Maier-Leibnitz Zentrum (MLZ), Germany\\}
	\altaffiliation{$^{\$}$ Deceased author\\}
	\email{$^{*}$ Email: shekhar@iitmandi.ac.in}	

	\begin{abstract}	
			
		The disordered pyrochlore oxide Dy$_{2}$Zr$_{2}$O$_{7}$ shows the signatures of field-induced spin freezing with remnant zero-point spin-ice entropy at 5 kOe magnetic field. We have performed zero-field and longitudinal field Muon spin relaxation ($\mu$SR) studies on Dy$_{2}$Zr$_{2}$O$_{7}$. Our zero field studies reveal the absence of both long-range ordering and spin freezing down to 62 mK. The $\mu$SR relaxation rate exhibits a temperature-independent plateau below 4 K, indicating a dynamic ground state of fluctuating spins similar to the well-known spin ice system Dy$_{2}$Ti$_{2}$O$_{7}$. The low-temperature spin fluctuations persist in the longitudinal field of 20 kOe as well and show unusual field dependence of the relaxation rate, which is uncommon for a spin-liquid system. Our results, combined with the previous studies do not show any evidence of spin ice or spin glass ground state, rather point to a disorder-induced dynamic magnetic ground state in the  Dy$_{2}$Zr$_{2}$O$_{7}$ material. 
		
	\end{abstract}
	
	\maketitle
	
	\section{Introduction}
	
In geometrically frustrated magnetic systems, the competing magnetic interactions combined with the lattice geometry prevent the system from establishing the long-range order. The corner-sharing tetrahedral arrangements of the magnetic rare earth ions in the pyrochlore lattice R$_{2}$B$_{2}$O$_{7}$ (R = rare-earth, B = transition metal) manifest the development of exotic magnetic states, which have been extensively studied in the literature \cite{bramwell2001spin,matsuhira2001novel,rau2019frustrated}. Variations at the R site can modulate the coherent magnetic ground state from a classical spin ice (Ho$_{2}$Ti$_{2} $O$_{7}$, Dy$_{2}$Ti$_{2}$O$_{7}$) to a quantum spin ice (Ce$_{2} $Zr$_{2}$O$_{7}$ and Nd$_{2}$Zr$_{7}$O$_{7}$), which can be attributed to rather different crystal-field ground states  of R sites in the D$_{3d}$ local symmetry. \cite{castelnovo2008magnetic,petit2016observation,gaudet2019quantum}. However, when structural disorder is introduced, the physical properties of the R ions in frustrated magnetic systems are strongly influenced. Furthermore, recent theoretical and experimental research probing disorder has provided interesting insight on the spin ices, spin liquids and XY pyrochlores  \cite{martin2017disorder,ross2012neutron,sala2014vacancy}. The emergence of disorder-induced magnetic states in frustrated magnets is one of the most important and current aspects of magnetism. The disorder often acts as a significant perturbation force that deviates a spin system far away from its coherent quantum ground state. This is particularly true for spin ice pyrochlore oxides R$_{2}$B$_{2}$O$_{7}$, in which disorder effects, such as stuffing and oxygen non-stoichiometry \cite{ross2012lightly,bowman2019role,sala2014vacancy}, or simply structural distortions \cite{martin2017disorder,sheetal2022structure}, or disorder due to the mixed B-site \cite{gomez2021absence}, are found to have a propounding impact on their magnetic ground state.\\

For instance,  Dy$_{2}$Ti$_{2}$O$_{7}$ is a clean pyrochlore spin ice system, whereas Dy$_{2}$Zr$_{2}$O$_{7}$ forms in a chemically disordered structural phase and develops a dynamic ground state down to 50 mK without zero-point entropy \cite{ramon2019absence}. Similar behavior is reported for another spin-ice material Ho$_{2}$Ti$_{2}$O$_{7}$, where the chemical alteration of the pyrochlore phase precludes the development of spin-ice correlations in Ho$_{2}$Zr$_{2}$O$_{7}$ \cite{sheetal2022field}. The Tb$_{2}$Ti$_{2}$O$_{7}$ remains in a dynamic magnetic state \cite{enjalran2004spin,takatsu2011quantum} whereas a subtle disorder in the lattice induces spin-glass behavior in Tb$_{2}$Hf$_{2}$O$_{7}$ \cite{sibille2017coulomb}. Theoretical studies show that varying the degree of disorder can cause a system to transition to different magnetic states, providing a better understanding of the physics of these structurally disordered frustrated systems \cite{savary2017disorder}. While a significant chemical disorder would likely preclude the development of spin ice correlation \cite{ramon2019absence,sheetal2022field}, a subtly tuned level of disorder may lead to a quantum spin liquid (QSL) ground state \cite{rau2019frustrated,savary2017disorder}. In a similar manner, it has been found that the disorder due to the site mixture of Mg$^{2+}$/Ga$^{3+}$ in the recently discovered triangular antiferromagnet YbMgGaO$_{4}$ also plays an important role in the formation of the QSL ground state in this compound \cite{zhu2017disorder,rao2021survival}. These studies show how disorder can lead to a plethora of intriguing phenomena in highly frustrated systems. The tuning of disorder offers an intriguing opportunity to investigate much new physics in frustrated magnets.\\

Along this line, Dy$_{2}$Zr$_{2}$O$_{7}$ has shown the possible realization of field-induced spin ice state by stabilizing the finite residual entropy of R[ln2-(1/2)ln(3/2)] at 5 kOe magnetic field, equivalent to Dy$_{2}$Ti$_{2}$O$_{7}$ \cite{sheetal2020emergence}. Unlike the spin ice Dy$_{2}$Ti$_{2}$O$_{7}$, the zero field heat capacity and neutron studies by Ramon \textit{et al.} suggests the highly dynamic spin state without any residual entropy down to 50 mK \cite{ramon2019absence}. Additionally, the ac susceptibility measurements show the signature of weak spin freezing at $\sim$1 K in the highly dynamic ground state \cite{ramon2019absence}. The neutron diffraction study in the presence of external magnetic field by the same group reveals the development of magnetic peak for \textit{H} $\geq$ 2 kOe, suggesting the long-range order state \cite{ramon2020geometrically}. This result does not conform with other macroscopic measurements and thus complicates the understanding of the origin of observed behavior and true ground state of Dy$_{2}$Zr$_{2}$O$_{7}$. Therefore, we have carried out zero-field and longitudinal field muon spin relaxation ($\mu$SR) measurements to understand the impact of the disorder on the low-temperature magnetic state and to determine whether or not the magnetic field transits the system to spin ice or the magnetically long-range ordered state.
	
	\section{Experimental techniques}
	
Polycrystalline Dy$_{2}$Zr$_{2}$O$_{7}$ was prepared by solid-state reaction method as described previously \cite{sheetal2021evolution}. The phase purity and crystal structure was characterized by performing the Rietveld refinement of the powder X-ray diffraction data. The obtained structural parameters were in close agreement with the previously reported parameters for the compound and confirmed the formation of the disordered pyrochlore phase. The $\mu$SR data was collected at the ISIS pulsed muon facility, Rutherford Appleton Laboratory, United Kingdom \cite{lord2011design} using HiFi spectrometer in the temperature range 62 mK and 4 K using a dilution refrigerator at various magnetic fields ranging between 0 - 20 kOe. Additionally, the high-temperature data up to 300 K were collected by transferring the sample to $^{4}$He cryostat.
	
	\section{Results and discusssion}
	
	\begin{figure}[tbp]
		\begin{center}
			\includegraphics[scale=0.55]{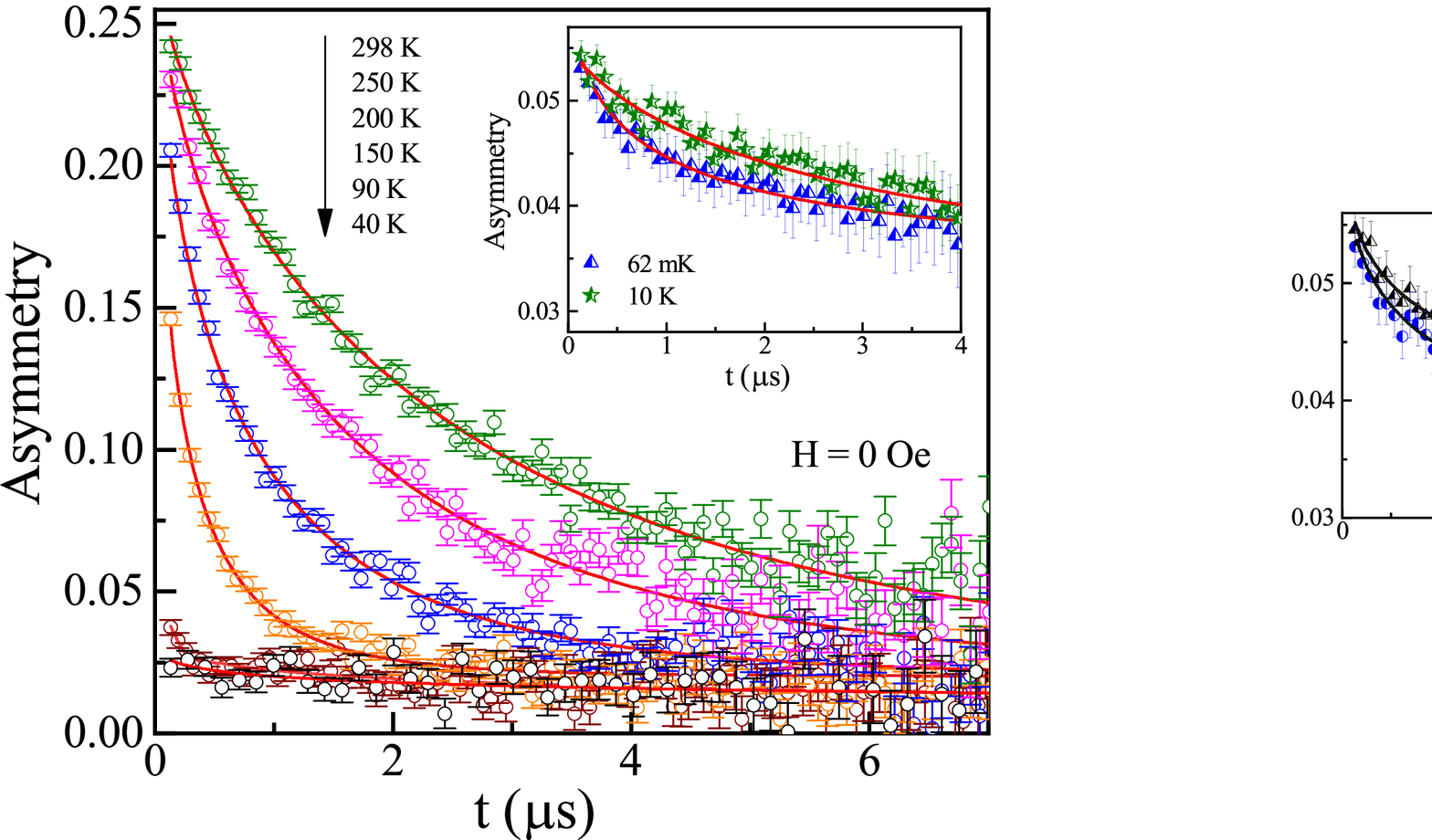}
			\caption{High-temperature zero-field $\mu$SR spectra for Dy$_{2}$Zr$_{2}$O$_{7}$. Solid (red) lines are the fit to an stretched exponential function described in the main text. Inset shows the relaxation spectra at \textit{T} = 62 mK and 10 K.} \label{fig:fig1}
		\end{center}
		\vspace{-10pt}
	\end{figure} 
	
	Fig. \ref{fig:fig1} show the selected zero-field $\mu$SR asymmetry spectra in the temperature range 62 mK and 300 K. The initial asymmetry shows a value of $\sim$ 0.25 at \textit{T} = 300 K and relaxes with time owing to the rapid fluctuations of the internal field of Dy$^{3+}$ moments in the paramagnetic phase. The relaxation signal decreases continuously on lowering the temperatures, and does not show oscillations within the analyzed time window of 0.12 to 15 $\mu$s. As expected and seen in the bulk magnetization studies, this behavior rules out the presence of long-range magnetic ordering and the formation of the static uniform local field. We have fitted the $\mu$SR asymmetry data with the stretched exponential function \textit{A(t)} = \textit{A$_{o}$}$exp[-(\lambda t)^{\beta}]$ + \textit{A$_{bg}$}, where A$_{o}$ is the initial asymmetry, $\lambda$ is the spin relaxation rate and $\beta$ is the stretched exponent. For the systems with a single well-defined spin fluctuation rate, the exponent $\beta$ = 1 and relaxation fits to the simple exponential function \cite{uemura1985muon}. However, for Dy$_{2}$Zr$_{2}$O$_{7}$, a simple exponential function was inadequate to account for the observed asymmetry spectra, therefore we used the phenomenological stretched exponential function. This indicates the presence of multiple spin fluctuation rates or relaxation channels, which is not surprising, given the structurally disordered pyrochlore lattice. This behavior resembles with the Tb$_{2}$Hf$_{2}$O$_{7}$ system exhibiting multiple relaxation rates due to the presence of anion-disordered pyrochlore lattice \cite{sibille2017coulomb}. \\
	
\begin{figure}[tbp]
		\begin{center}
			\includegraphics[width=8.5cm,height=6.5cm]{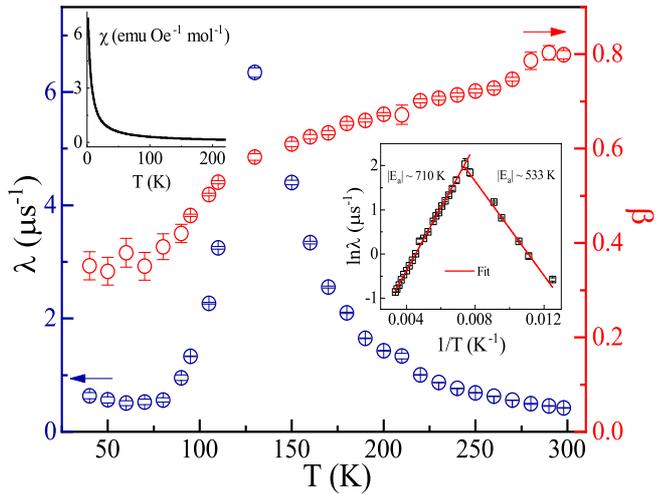}
			\caption{Temperature dependence of the parameters $\lambda$ and $\beta$ extracted from the stretched exponential fit of the zero field muon spin relaxation data collected between 40 - 300 K. The standard deviation of the parameters is represented by the error bars. The maximum at $\sim$130 K suggests the relaxation rate going beyond the limit of the instrument as a result of large moment value of Dy$^{3+}$ ion. Inset (left) DC susceptibility of Dy$_{2}$Zr$_{2}$O$_{7}$ taken from Ref\cite{sheetal2020emergence} does not show any long-range magnetic ordering and (right) shows the Arrhenius fit of the relaxation rate for \textit{T} $\geq$ 40 K and \textit{T} $\geq$ 130 K. } \label{fig:fig2}
		\end{center} 	 \vspace{-25pt}
	\end{figure}

	The temperature dependence of $\lambda$ and $\beta$ of stretched function obtained from the time dependent muon asymmetry data are plotted in Fig. \ref{fig:fig2}. As seen in the figure, spin relaxation rate $\lambda$ increases and the exponent $\beta$ decreases on cooling. This is due to the cumulative effect of the depopulation of Dy$^{3+}$ excited crystal field levels and slowing down of spin fluctuations leading to a wider field distribution at muon sites. The relaxation rate for Dy$_{2}$Zr$_{2}$O$_{7}$ peaks around \textit{T} $\sim$ 130 K, showing similarities with Dy$_{2}$Ti$_{2}$O$_{7}$ system \cite{dunsiger2011spin,lago2007musr}. Such a peak in $\lambda$(T) is a characteristic feature of the changes in spin dynamics, which is often associated with the magnetic transition \cite{goko2017restoration}. While being consistent with the earlier $\mu$SR studies on powder Dy$_{2}$Ti$_{2}$O$_{7}$, this feature suggests a crossover to the slow fluctuating regime in Dy$_{2}$Zr$_{2}$O$_{7}$ without long-range ordering \cite{dunsiger2011spin}. The DC susceptibility measurements (shown in the inset of Fig. \ref{fig:fig2}) also suggest the absence of any long-range magnetic ordering near this temperature and also down to 2 K, indicating the paramagnetic ground state \cite{sheetal2020emergence}.\\
	
\begin{figure}[tbp]
		\begin{center}
			\includegraphics[width=8.5cm,height=6.5cm]{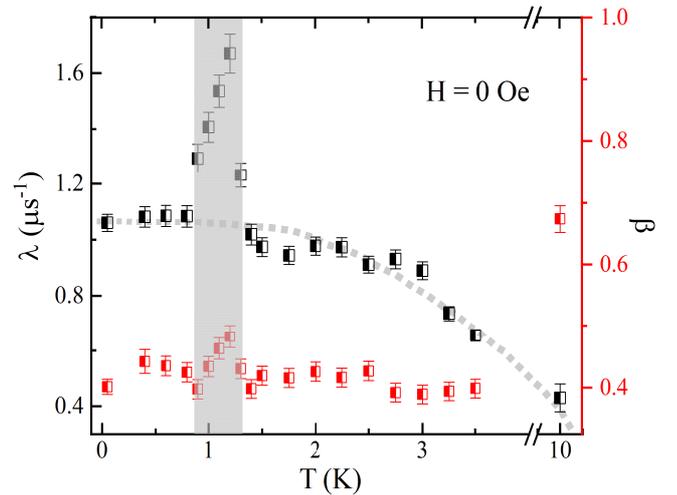}
			\caption{Temperature dependence of the parameters $\lambda$ and $\beta$ extracted from the stretched exponential fit of the zero field muon spin relaxation data collected between 0.062 - 10 K. The standard deviation of the parameters is represented by the error bars. A peak shape anomaly at $\sim$1.2 K (in the shaded region) is due to the effect of variation of stray field due to field ramp on other nearby instrument for these data points.} \label{fig:fig3} 
		\end{center}  \vspace{-10pt}
	\end{figure}

	The inset of Fig. \ref{fig:fig2} shows the Arrhenius fit of extracted relaxation rate to $\lambda$ $\propto$ exp($E_{a}/k_{B}T$), yielding an activation energy of $E_{a}/k_{B}$ $\sim$ 533 K for \textit{T} $\geq$ 40 K and $E_{a}/k_{B}$ $\sim$ 710 K for \textit{T} $\geq$ 130 K. Note that the Arrhenius fits to the relaxation rate in Dy$_{2}$Ti$_{2}$O$_{7}$ yield an energy barrier of 468 K for \textit{T} $\geq$ 60 K \cite{lago2007musr} and a barrier of 210 K for high-temperature spin freezing at $\sim$ 16 K \cite{matsuhira2001novel}. This indicates the thermally activated spin dynamics in the high-temperature regime. No activation barrier could be calculated for 10 K $\leq$ \textit{T} $\leq$ 40 K region, as the polarization signal got completely lost below \textit{t} $<<$ 1 $\mu$s. However, on further cooling below 10 K, a recovery of the resolved asymmetry at \textit{t} = 0.12 $\mu$s, (with comparatively less than 1/3 value of initial asymmetry value as expected for spin glass-like freezing) was observed (see inset of Fig. \ref{fig:fig1}). Generally, for a magnetic system, the re-polarization of muon signal is a signature of the emergence of static internal fields associated with the spin freezing state. However, considering the earlier studies of heat capacity and diffuse neutron scattering, re-polarization of muon signal for Dy$_{2}$Zr$_{2}$O$_{7}$ suggests the development of the fluctuating correlated spin state \cite{ramon2019absence}. Our preliminary measurements were also limited to 0.12 $\leq$ \textit{t} $\leq$ 15 $\mu$s and it may possible that the information about the low-temperature static spin state (if any) is lost in the missing time window of \textit{t} $<$ 0.12 $\mu$s, as expected for the glassy/freezing state. Therefore, $\mu$SR measurements in the early time windows would be useful for ascertaining the magnetic ground state of Dy$_{2}$Zr$_{2}$O$_{7}$. The stretched exponent $\beta$ gradually decreases from $\sim$ 0.8 at high temperatures to a constant value of $\sim$ 0.4 (see Fig. \ref{fig:fig2} and \ref{fig:fig3}). In contrast, the spin-glass systems show a dramatic drop in $\beta$ nearly equaling 1/3 as the system approaches freezing temperature \cite{campbell1994dynamics}. \\
	
	Figure \ref{fig:fig3} shows that the relaxation rate increases below 10 K and nearly saturates below 3 K without any signature of the non-relaxing tail. This demonstrates the absence of further slowing-down of the spin dynamics as observed in the spin-ice materials. We observed a substantial $\lambda$ $\sim$ 1 $\mu$s$^{-1}$ relaxation rate at \textit{T} $\leq$ 3 K. This behavior is similar to Dy$_{2}$Ti$_{2}$O$_{7}$, however, unexpected for spin ice system of Ising spins where the monopole excitations are separated by large energy barriers against single spin-flip process \cite{jaubert2009signature}. This suggests that the spin correlations observed in Ref\cite{ramon2019absence} in diffuse neutron scattering at low temperatures fluctuate on a different time scale from the muon Larmor precession frequency (MHz). Although a similar persistent spin dynamics have been observed in a variety of frustrated magnets, the origin of the temperature-independent relaxation plateau remains unknown \cite{mcclarty2011calculation,uemura1994spin}. On the other hand, quantum spin liquid systems investigated using ZF-$\mu$SR also exhibit temperature independent muon relaxation rate at the lowest temperature range, and then follows a power law behavior with increasing temperature \cite{yueschengPRL,ABhattacharyaPRB}.  It is to mention that Ramon \textit{et al.} found a strong frequency dependence in Dy$_{2}$Zr$_{2}$O$_{7}$ below 2 K in ac susceptibility studies \cite{ramon2019absence}, implying the presence of unusual spin freezing as seen in spin ice Dy$_{2}$Ti$_{2}$O$_{7}$, Ho$_{2}$Ti$_{2}$O$_{7}$ and other frustrated magnets. However, no such feature is observed in our $\mu$SR data, usually distinguished by a peak shape anomaly at the freezing temperature \cite{dunsiger1996muon}. Upon the recovery of asymmetry, the relaxation rate increases by 60$\%$ when cooling from T $>$ 9 K (Curie-Weiss temperature) down to 0.062 K, indicating a weak slowing down of the Dy$^{3+}$ spin fluctuations. Whereas, below the freezing point, relaxation rate in a spin-glass typically increases by several orders of magnitude ($\lambda$ $\sim$ 1–20 $\mu$s$^{-1}$) \cite{uemura1985muon,bono2004mu}. Instead, the presence of finite residual spin relaxation rate down to the lowest measuring temperature is consistent with the large entropy of Rln2 \cite{sheetal2020emergence} and large susceptibility value \cite{ramon2019absence}, providing evidences of magnetic excitation in a weakly frozen spin state. \\
	
	\begin{figure}[tbp]
		\begin{center}
		\hspace{-0.4cm}	\includegraphics[scale=0.5]{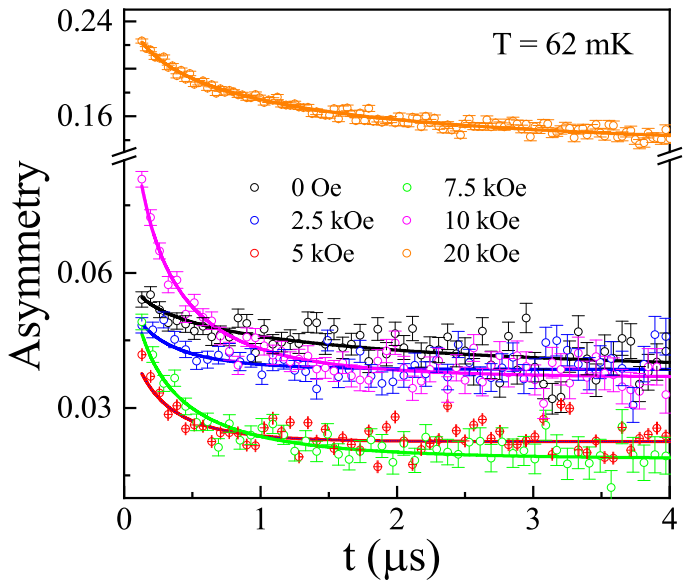}
			\caption{Muon spin relaxation spectra collected at \textit{T} = 62 mK in a longitudinal field of 0 - 20 kOe. Color line represents the fit using stretched exponential function (as described in the main text).}  \label{fig:f1}
		\end{center}	 \vspace{-20pt}
	\end{figure}  
		
\begin{figure}[tbp]
		\begin{center}
		\hspace{-0.5cm}	\includegraphics[scale=0.5]{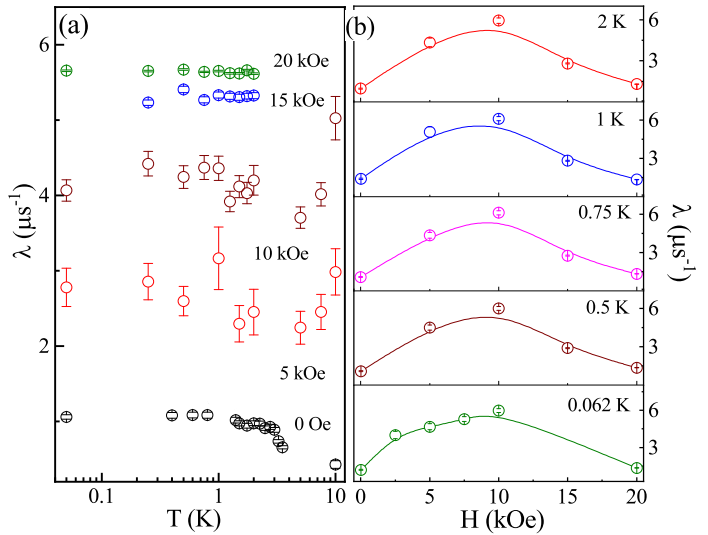}
			\caption{ Temperature (a) and field (b) dependence of longitudinal relaxation rate of Dy$_{2}$Zr$_{2}$O$_{7}$. The $\lambda$ values for 15 and 20 kOe field in (a) are shifted upwards in y-axis (by $\sim$ 4 $\mu$s$^{-1}$) for better clarity. The stray fields contribution in zero-field data is removed for the sake of comparison.} \label{fig:f2}
		\end{center}  \vspace{-10pt}
	\end{figure}

       Fig. \ref{fig:f1} shows the asymmetry versus time spectra collected in the longitudinal fields of \textit{H} = 0 - 20 kOe. Similar to  zero field measurement, $\mu$SR spectra in the magnetic field do not exhibit any evidence of magnetic ordering down to 62 mK even up to the applied field of 20 kOe. These $\mu$SR spectra taken at different fields are also fitted well by the stretched exponential function. The obatined temperature and magnetic field dependence of the extracted relaxation rate are plotted in Fig. \ref{fig:f2}(a) and \ref{fig:f2}(b) respectively. The relaxation rate has a temperature dependence similar to the zero-field for all fields. Although the Dy$^{3+}$ moments undergo a transition to the field-induced frozen state on the ac susceptibility time scale below 10 K, fluctuations appear to emerge in longitudinal-field muon experiments down to 62 mK. In comparison to muon technique which probes the spin dynamics of the order of 10$^{4}$ to 10$^{11}$ Hz, ac susceptibility employed for this study could probe the fluctuations down to the millisecond timescale (or kHz) only \cite{blundell2022muon}. \\
	
	The longitudinal relaxation rate as a function of the applied field, for different temperatures (\textit{T} = 0.062 - 2 K) shows a peak around 10 kOe, indicating a change in spin dynamics. This feature reflects the competition between partially frozen (static) and dynamic components of the local field at the muon site. Further, the field dependence of the relaxation rate and the higher time tail observed in the $\mu$SR (Fig. 4), are due to the field dependence of the fast fluctuating electronic relaxation and it persists up to the highest field of 20 kOe. This confirms the paramagnetic correlated dynamical ground state without long range ordering uo to 20 kOe at 62 mK.  This behavior is consistent with the field-induced relaxation in ac susceptibility \cite{sheetal2020emergence}. Though the spin-ice entropy is reported to stabilize in Dy$_{2}$Zr$_{2}$O$_{7}$ at \textit{T} = 5 kOe below 1 K, the longitudinal field $\mu$SR results do not show any supported signature, rather suggest the development of unusual field-induced magnetic state in a highly dynamic background of fluctuating spins. This behavior is similar to other spin ice systems, where $\mu$SR reveals a large density of excitation in the two-in-two-out frozen spin state \cite{lago2007musr}. \\
	
	In conclusion, there are no evidences of long-range ordering and spin freezing down to 62 mK  in our comprehensive studies of ZF and LF $\mu$SR. In spite of the weak static field as shown by ac susceptibility and also evident from the recovery of asymmetry below 10 K, Dy$_{2}$Zr$_{2}$O$_{7}$ exhibits the persistence of dynamic magnetism down to 62 mK. Though the absence of long-range ordering, persistent spin fluctuations and saturation of $\lambda$ at low temperatures are the signature of quantum spin liquid ground state (QSL), the unusual dependence of the relaxation rate on field exclude this possibility and imply the development of magnetically correlated spin state. It is to mention here that, neutron diffraction measurement by Ramon \textit{et. al.} \cite{ramon2020geometrically} shows the emergence of magneitc Braggs peaks for \textit{H} $\geq$ 2 kOe, however the broad short-range correlation peak remain centered at 1.2 $\AA^{-1}$ even at the highest applied magnetic field. This could indicate the coexistence of a dynamic spin state and a partially ordered spin state. In $mu$SR, however, the homogeneous field cannot be seen in the presence of fluctuations. The observed peculiar behavior is might be an  outcome of the presence of the lattice disorder in comparison to structurally clean pyrochlore Dy$_{2}$Ti$_{2}$O$_{7}$ system. The structural studies Dy$_{2}$Zr$_{2}$O$_{7}$ in Ref\cite{ramon2020geometrically} also suggests the site mixing of Dy/Zr ions results bond randomness in the lattice. Similarly, YbMgGaO$_{4}$ possess QSL gorund state attributed to the presence of Mg/Ga site mixing \cite{zhu2017disorder}. Consequently, the dominant effect from A/B site mixing or randomness along with magnetic frustration is responsible for the curbing of ordering and the dynamic ground state.  \\
			
	\noindent	
	\textbf{Acknowledgment:} \\
	We thanks AMRC, Indian Institute of Technology Mandi for the experimental facility. Sheetal Acknowledged IIT Mandi and MHRD India for the HTRA fellowship. Dr. D. T. Adroja thank EPSRC UK for funding (Grant No. EP/W00562X/1). We would like to thank the ISIS Facility  for beam time, RB2010601 and the data are available from Ref\cite{sunravelling}.
	
	\bibliography{Ref}
	
\end{document}